\documentclass[twocolumn]{svjour3}         
\smartqed  

\usepackage{graphics}
\usepackage{graphicx}
\usepackage{epstopdf}
\def\cms{cm$^{-1}\,$}
\def\cm{cm$^{-1}$}
\def\Szo{S$_0$$\rightarrow$S$_1\,$}

\begin{document}

\title{Laser control of double proton transfer in porphycenes}
\subtitle{Towards an ultrafast switch for photonic molecular wires}


\author{Mahmoud K. Abdel-Latif \and Oliver K\"uhn
}
\institute{M. K. Abdel-Latif \at
              Institut f\"ur Physik, Universit\"at Rostock, D-18051 Rostock, Germany
              and 
              Chemistry Department, Faculty of Science, Beni-Suef University, Beni-Suef, Egypt
           \and
           O. K\"uhn \at
              Institut f\"ur Physik, Universit\"at Rostock, D-18051 Rostock, Germany\\
              Tel.: +49-381-498-6950\\
              Fax:  +49-381-498-6942\\
              \email{oliver.kuehn@uni-rostock.de}           
}

\date{\today}
\maketitle
\begin{abstract}
Electronic excitation energy transfer along a molecular wire depends on the relative orientation of the electronic transition dipole moments of neighboring chromophores. In porphycenes this orientation is changed upon  double proton transfer in the electronic ground state. We explore the possibility to trigger such a double proton transfer reaction by means of an infrared pump-dump laser control scheme. To this end a quantum chemical characterization of an asymmetrically substituted porphycene is performed using density functional theory. Ground state geometries, the topology of the potential energy surface for double proton transfer, and \Szo transition energies are compared with the parent compound porphycene and a symmetric derivative. Employing a simple two-dimensional model for the double proton transfer, which incorporates sequential and concerted motions, quantum dynamics simulations of the laser driven dynamics are performed which demonstrate tautomerization control. Based on the orientation of the transition dipole moments this tautomerization may lead to an estimated  change in the F\"orster transfer coupling of about 60\%.
\end{abstract}
\section{Introduction}
\label{intro}
Molecular photonic wires are of considerable  interest as biomi\-me\-tic models for photosynthetic light harvesting systems and as molecular devices \cite{balzani03}.  Lindsey et. al. have been the first to demonstrate directed energy transfer through a porphyrin-based wire after selective excitation of a boron-dipyrrin donor \cite{wagner94:9759}. Subsequently, they described a number of different systems on the basis of multiporphyrin arrays \cite{wagner96:11166,hsiao96:11181,seth96:11194} including, for instance, shape persistent cyclic architectures \cite{li99:8927} or optoelectronic gates  \cite{ambroise01:1023,lammi01:5341} (for a review, see also Ref.  \cite{holten02:57}). The energy transfer in these structures is dominated by a through-bond mechanism with some dependence on the nature of the bridge linking the chromophores \cite{song09:16483}.
A different design strategy has been followed by Sauer and coworkers who have used the backbone of double-stranded DNA as a scaffold for positioning highly fluorescent chromophores \cite{heilemann04:6514,tinnefeld05:217,sanchez-mosteiro06:26349,heilemann06:16864} (see also Ref. \cite{vyawahare04:1035}). Here, F\"orster type dipole-dipole interaction driven transfer is the responsible mechanism for energy transfer with efficiencies close to 100\% over a range of $\sim$14 nm.

The utilization of molecular wires in optoelectronic devices requires to have at hand a means for switching the energy transfer dynamics. Various methods such as  optical excitation using UV radiation \cite{bahr01:7124},  electrochemical oxidation \cite{lammi01:5341,wagner96:3996,akasaka02:130} or changing the pH value of the medium \cite {albelda02:63} have been suggested  (for an overview, see also Refs. \cite{balzani03,otsuki08:32}). This includes even the realization of photochemical logic gates as reported in Ref. \cite{straight07:777}. Recently, we have proposed another route to switch excitation energy transfer which could operate in the ultrafast regime of a few picoseconds \cite{abdel-latif10:76}. The idea builds on the observation that the direction of the electronic transition dipole moment, which is responsible for the coupling between chromophores in the F\"orster mechanism \cite{may10}, depends on the positions of the two  H atoms in the porphyrin like building blocks of molecular wires. Thus switching between the ground state tautomers, i.e. triggering the double proton transfer (DPT) by means of an appropriately designed laser field in the infrared (IR) domain enables one to modify the strength of the F\"orster transfer coupling.

The aim of the present contribution is to extend the model study of Ref. \cite{abdel-latif10:76} towards a specific molecular system and to show exemplarily to what extent laser-triggered DPT can influence the F\"orster transfer coupling. 
Specifically, we will focus on porphycene like structures, which have rather strong hydrogen bonds (HBs) whose properties are widely adjustable by the structural design \cite{waluk07:245}. For the laser-driven DPT the shape of the potential energy landscape is rather important. Besides energy barriers which need to be surmounted by excitation with a few IR photons, reactants and products should be energetically different to prevent an efficient backreaction via tunneling. Asymmetrically substituted porphycenes seem to meet these requirements. Below we will consider 9-actetoxy-2,7,12,17-tetra-n-pro\-pyl\-porphycene (\textbf{3}) whose spectroscopy  had been studied by Waluk and coworkers \cite{gil00:534}. Thus, besides the specific goal of demonstrating a laser-driven switch for molecular photonic wires, our quantum chemical studies of the ground state properties and the {\Szo} ex\-ci\-tation spectrum will shed some light on the effect of symmetric (\textbf{1}) and asymmetric (\textbf{3}) substitution of the porphycene parent compound (\textbf{1}) and hence will provide the theoretical basis for the assignment of the absorption spectrum.

\begin{figure}[h!]
\begin{center}
\resizebox{0.8\columnwidth}{!}{
\includegraphics{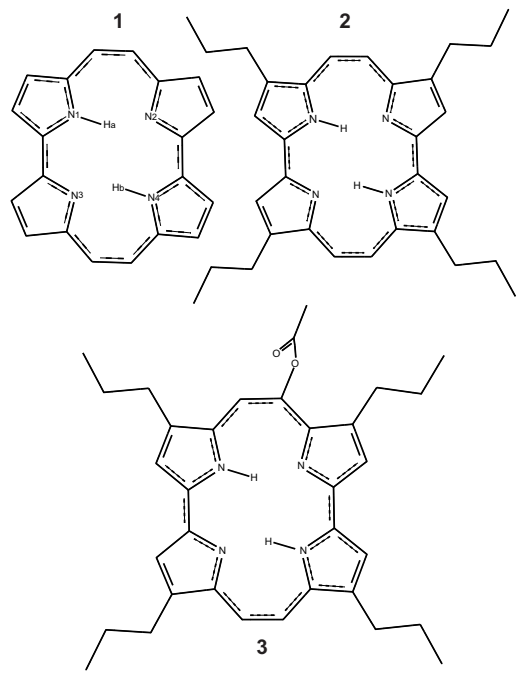}
}
\end{center}
\caption{Reference structures of considered porphycenes: \textbf{1} free base porphycene, \textbf{2}  tetra-n-propyleporphycene,  and \textbf{3} more stable trans form (Tr1) of tetra-n-propyleacetoxyporphycene.}
\label{fig:1}     
\end{figure}

The paper is organized as follows: Section \ref{sec:methods} gives a brief summary of the quantum chemical and quantum dynamical methods. The results Section \ref{sec:res} starts with the discussion of the geometries and energetics of the stationary points of porphycene and its derivatives (see Fig. \ref{fig:1}).  Based on this information a two-dimensional (2D) model Hamiltonian is constructed in Section \ref{sec:ham} which comprises coordinates of concerted and sequential DPT.  Subsequently,  the \Szo excitation spectrum is discussed in terms of the involved molecular orbitals and transition dipole moments. The laser driven DPT dynamics will be presented in Section \ref{sec:control} and the final Section \ref{sec:concl} concludes with a summary.
%
%
%
\begin{figure}[t]
\begin{center}
\resizebox{1.0\columnwidth}{!}{
\includegraphics{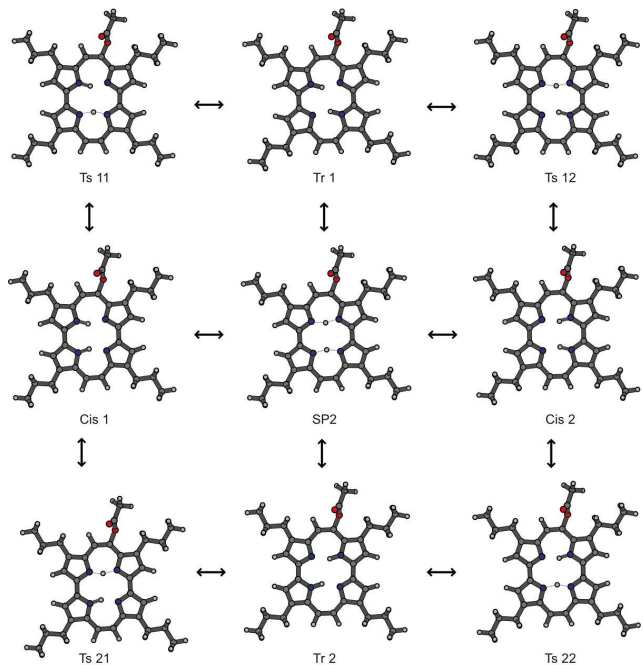}
}
\end{center}
\caption{Optimized geometries (B3LYP/6-31+G(d,p)) of all stationary points of \textbf{3} which are relevant for DPT.}
\label{fig:3struct}     
\end{figure} 
%

\section{Methods}
\label{sec:methods}
Geometry optimization for the electronic ground state has been performed using density functional theory (DFT) employing the B3LYP hybrid functional together with a 6-31+G(d,p) basis set. Stationary points have been validated by means of frequency calculations. Excitation energies are obtained from time-dependent DFT (TDDFT) calculations at the optimized ground state geometries (vertical excitation) and using the same functional and basis set. All quantum chemical calculations are performed with the Gaussian 03 suite of programs \cite{g03}.

The two-dimensional time-dependent Schr\"odinger e\-qua\-ti\-on has been solved using the multiconfiguration time-dependent Hartree (MCTDH) approach \cite{meyer90:73,beck00:1} as implemented in the Heidelberg program package \cite{mctdh84}. The grid was chosen according to the  harmonic oscillator discrete variable representation (64 points with\-in [-2.5:2.5]a$_{\rm 0}$).  The actual propagation is performed using the variable mean field scheme  in combination with  a 6-th order  Adams-Bashforth-Moulton integrator. Using 20 single particle functions per degree of freedom the largest natural orbital populations have been typically on the order of $10^{-6}$. Selected eigenstates of the time-inde\-pen\-dent Hamiltonian, $\varphi_i$, are obtained by improved relaxation \cite{meyer06:179}.
%
\section{Results}
\label{sec:res}
\subsection{Optimized geometries}
\label{sec:geom}
Ground state optimized structures of \textbf{2} show that this molecule like \textbf{1} is planar in accord with Ref. \cite{waluk07:245}. Asymmetric substitution, however, causes  \textbf{3} to become nonplanar with respect to the acetoxy group. Overall \textbf{3} has two global trans minima, Tr1 and Tr2, two local cis minima, Cis1 and Cis2, four local maxima which correspond to first order saddle points, Ts11 to Ts22, and one global maximum being a second order saddle point, SP2. All these structures are shown in Fig. \ref{fig:3struct} and the energetics is compiled in Tab. \ref{tab:energies}. In terms of the double proton transfer, SP2 is passed in the concerted mechanism, while pathways like Tr1-Ts11-Cis1-Ts21-Tr2 are followed during sequential transfer. Finally, we note from the dipole moments given in  Tab. \ref{tab:energies} that \textbf{3} is considerably more polar than  \textbf{2}.

The energetic difference between trans and cis, which is  804 \cms for \textbf{1}, is largely unaffected by the symmetric substitution in \textbf{2}. The same holds true for the second order transition state SP2, i.e. the barrier height for concerted DPT is about 2260 \cm. The barrier for stepwise proton transfer which is 1637 \cms for \textbf{1} decreases slightly to 1514 \cms in \textbf{2}. Upon asymmetric substitution with an acetoxy group in the 9 position in \textbf{3} most notably in comparison with \textbf{2} is the asymmetry of the two trans tautomers which amounts to 38 \cm. This rather small value is in good accord with the experimental finding of indistinguishable tautomers \cite{gil00:534}. Further, the SP2 transition state is lowered by about 4\% and the first order saddle points and cis minima are changed in a way such as to make a stepwise DPT via Cis1 energetically more preferable.

\begin{table} [tp]%
\caption{Energies and absolute value of dipole moments of stationary  
points for the tautomerism of the molecules shown in Fig. \ref{fig:1}  
at the B3LYP/6-31+G(d,p) level of theory (values for \textbf{1} from   
Ref. \cite{smedarchina07:314}).}
\begin{tabular}{|c|c|r|c|}
\hline
molecule & structure & $\Delta E$ (\cm)  & dipole (Debye)\\
\hline
\textbf{3} & Tr1 & 0& 1.633\\
&Tr2& 38& 1.481\\
&Cis1& 726& 2.042\\
&Cis2& 838 & 2.218\\
&SP2& 2169 & 1.635\\
&Ts11& 1329 & 1.539\\
&Ts12& 1505 & 1.865\\
&Ts21& 1479 & 1.690\\
&Ts22& 1461 & 1.834\\
\hline
\textbf{2} & Tr & 0 & 0.002\\
&Cis& 817 & 1.452\\
&SP2& 2263 & 0.000\\
&SP1& 1514 & 0.770\\
\hline
\textbf{1} & Tr & 0 &\\
&Cis& 804 &\\
&SP2& 2256 &\\
&SP1& 1637 &\\
\hline
\end{tabular}
\label{tab:energies}
\end{table}
%
%

The results on the HB geometries of \textbf{3} are compiled in Tab. \ref{tab:geo} which also contain values for the trans tautomers of \textbf{1} and \textbf{2}. Comparing the most stable trans forms of the three molecules the differences in HB parameters are rather small.  Most notably again is the asymmetry, i.e. in Tr1 the HB next to the acetoxy substitution site (N$_1$-H$_a$-N$_2$) is slightly shorter and substantially more nonlinear as compared with the 
N$_3$-H$_b$-N$_4$ HB. The geometric signature of the energetically more preferable path for stepwise DPT can be found in the stronger contraction of the N$_3$-H$_b$-N$_4$ HB in Ts11 as compared with the  N$_1$-H$_a$-N$_2$ HB in Ts12. The transition states for the second transfer step, Ts21 and Ts22, on the other hand, have almost the same HB parameters.

%
\begin{table*} [t]%
\centering
\caption{Geometries of the HBs in \textbf{3} (and for Tr in \textbf{1} and \textbf{2}) at the B3LYP/6-31+G(d,p) stationary points (bond lengths in \AA ngstrom, angles in degrees, data for \textbf{1} are taken from Ref.  \cite{shibl07:315}).}
\begin{tabular}[t]{|l|cccc|cccc|}
\hline
structure & N$_1$-H$_a$ & N$_2$-H$_a$ & N$_1$-N$_2$ &  $\angle$N$_1$-H$_a$-N$_2$ & N$_3$-H$_b$ & N$_4$-H$_b$ & N$_3$-N$_4$ &  $\angle$N$_3$-H$_b$-N$_4$ \\
\hline
Tr1         & 1.05 & 1.68 & 2.65 & 152.9 & 1.05 & 1.70 & 2.67 & 150.9 \\ 
Tr2         & 1.68 & 1.05 & 2.65 & 153.0 & 1.67 & 1.05 & 2.66 & 154.9 \\ 
Cis1       & 1.61 & 1.07 & 2.61 & 155.1 & 1.06 & 1.64 & 2.62 & 152.6 \\ 
Cis2       & 1.07 & 1.61 & 2.61 & 154.0 & 1.60 & 1.07 & 2.62 & 156.5 \\ 
SP2        & 1.30 & 1.26 & 2.51 & 159.4 & 1.28 & 1.27 & 2.50 & 159.1 \\ 
Ts11      & 1.06 & 1.61 & 2.60 & 153.1 & 1.26 & 1.30 & 2.52 & 159.2 \\ 
Ts12      & 1.30 & 1.26 & 2.52 & 159.6 & 1.57 & 1.07 & 2.58 & 154.8 \\ 
Ts21      & 1.26 & 1.31 & 2.52 & 159.2 & 1.07 & 1.57 & 2.58 & 155.3 \\ 
Ts22      & 1.57 & 1.07 & 2.59 & 156.8 & 1.31 & 1.25 & 2.52 & 158.9 \\ 
\hline
\textbf{2}            & 1.05 & 1.69 & 2.67 & 152.6  &&&&\\
\textbf{1} & 1.05 & 1.68 & 2.66 & 152.8 &&&& \\
\hline
\end{tabular}
\label{tab:geo}
\end{table*}
%

\subsection{Ground state model Hamiltonian}
\label{sec:ham}
The potential energy surface for DPT in the electronic ground state will be modeled using a simple Hamiltonian which was introduced  by Sme\-dar\-chi\-na and coworkers \cite{smedarchina07:174513,smedarchina08:1291}  and which has been supplemented by asymmetric terms to mimic systems like \textbf{3} in Ref. \cite{abdel-latif10:76}. Assuming that only the proton coordinates are active on a surface where all other coordinates are adiabatically adjusted we have the following Hamiltonian ($m=2 m_{\rm H}$)
\begin{eqnarray}
\label{eq:ham0}
H&=&-\frac{\hbar^2}{2 m}\left( \frac{\partial^2}{\partial x_s^2}+\frac{\partial^2}{\partial x_a^2}\right) + U_{\rm sym}(x_s,x_a) \nonumber\\
&+& U_{\rm asym}(x_s,x_a)\, .
\end{eqnarray}
Here, symmetric and asymmetric transfer coordinates $x_s$ and $x_a$, respectively, have been introduced. 
They are defined with respect the to linear displacements of the two H atoms along the hydrogen bonds, $x_1$ and $x_2$, as follows: $x_s=(x_1+x_2)/2$ and  $x_a= (x_1-x_2)/2$.
The symmetric potential, $U_{\rm sym}(x_s,x_a)$, originates from two bilinearly coupled quartic oscillators when expressed in the coordinates of the individual H atoms $x_1$ and $x_2$. After introduction of the transfer coordinates $x_s$ and $x_a$ this gives:
\begin{eqnarray}
\label{eq:U0as}
U_{\rm sym}(x_s,x_a) & = & 2U_0 + \frac{U_0}{x_0^2} 
\left[(g-4)x_a^2-(g+4)x_s^2\right] \nonumber \\
 & + & \frac{2U_0}{x_0^4}(x_s^4+x_a^4+6x_s^2x_a^2) \,.
\end{eqnarray}
Here $g$ is the coupling constant, $\pm x_0$ are the minima for the uncoupled double minimum potential which has a barrier height of $U_0$. Asymmetry can be introduced by the following term:
\begin{equation}
\label{eq:Uasym}
U_{\rm asym}(x_s,x_a)=\frac{\alpha_{\rm trans} U_0}{x_0} x_s + \frac{\alpha_{\rm cis} U_0}{x_0}x_a 
\end{equation}
where $\alpha_{\rm trans}$ and $\alpha_{\rm cis}$ are dimensionless parameters characterizing the detuning between the trans and cis states, respectively.

\begin{figure}[h!]
\begin{center}
\resizebox{0.9\columnwidth}{!}{
\includegraphics{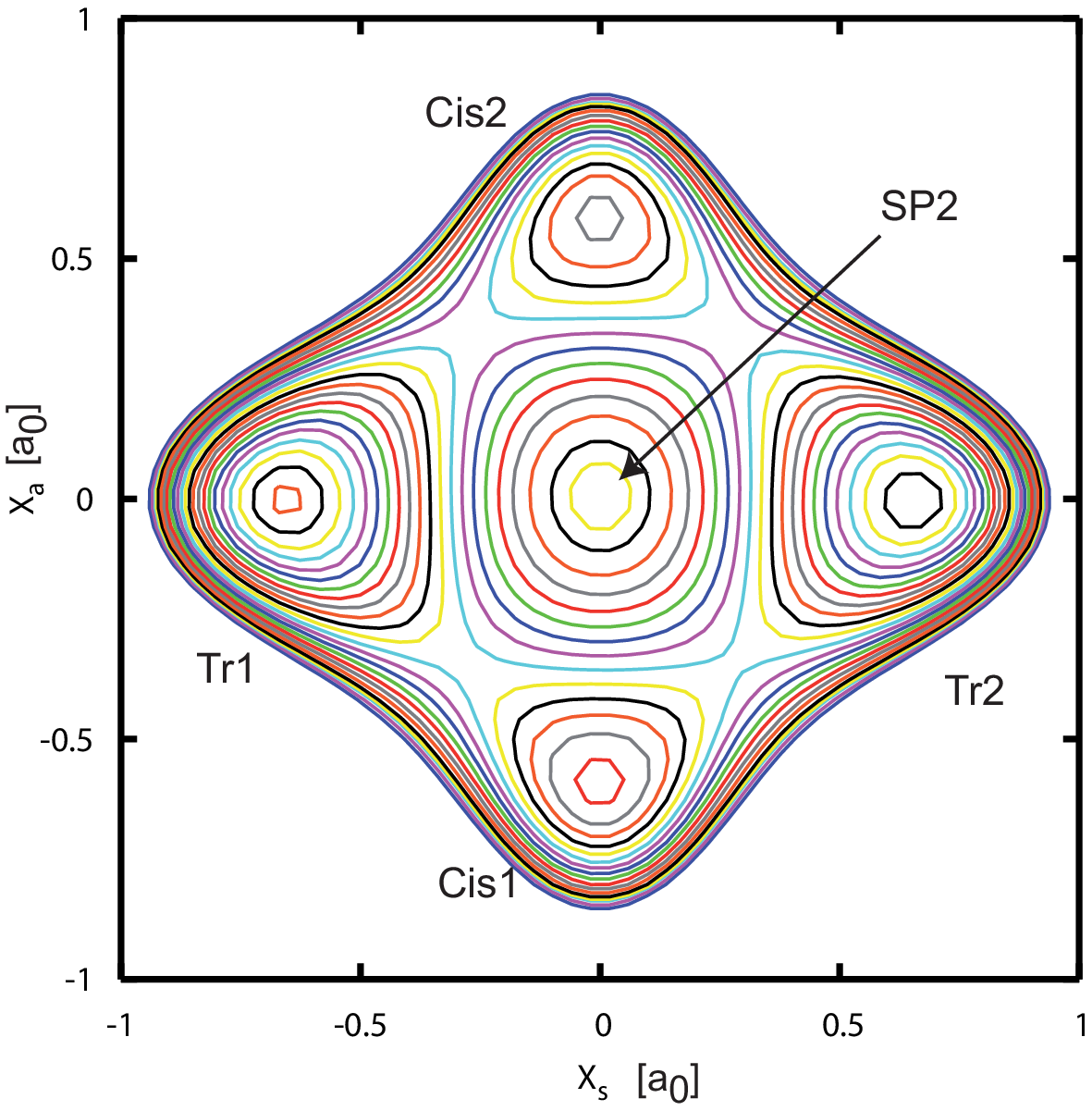}
}
\end{center}
\caption{Two-dimensional model potential according to Eq. (\ref{eq:ham0}) for the DPT in \textbf{3}. The parameters are $U_0=876$ cm$^{-1}$, $x_0=0.624$a$_0$, $g=0.435$, $\alpha_{\rm trans}=0.0095$, and $\alpha_{\rm cis}=0.034$. The  barrier height for concerted transfer is 2169 cm$^{-1}$. The energetic difference between the two trans minima is 38 cm$^{-1}$ and the  energies of the cis minima are 725 \cms and 838 cm$^{-1}$ (contour lines are in steps of 400 \cms from 400 \cms to 8800 \cm).}
\label{fig:pes}     
\end{figure}

The parameters  for this Hamiltonian have been obtained by fitting to the quantum chemical results of Tab. \ref{tab:energies}. The resulting potential energy surface is plotted in Fig. \ref{fig:pes} and the fitting parameters are given in the figure caption. Note that while the minima and the SP2 maximum are in good agreement with the quantum chemistry results, the simple form of the potential does not provide enough flexibility for obtaining all first order  transition states at the same time with comparable accuracy. The deviations range from 51  \cms to 169 \cm, which is, however, still acceptable given the overall approximate nature of the treatment.

\begin{figure}[t]
\begin{center}
\resizebox{0.9\columnwidth}{!}{
\includegraphics{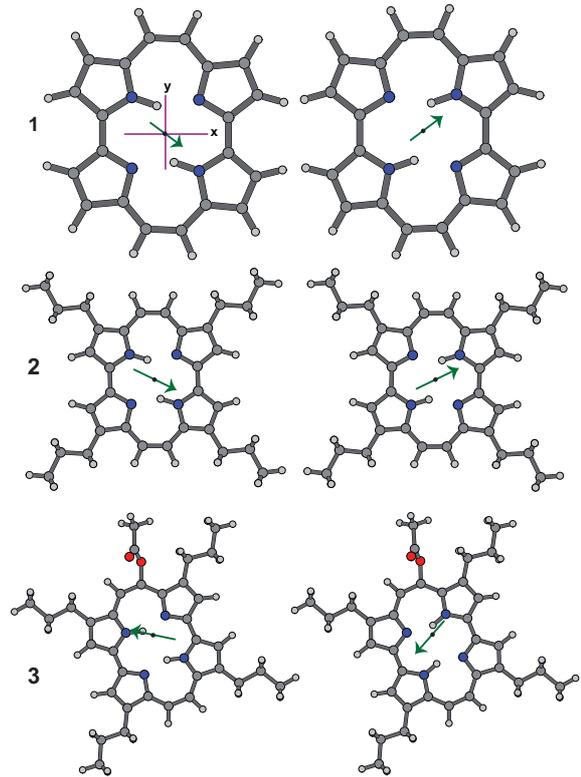}
}
\end{center}
\caption{Transition dipole moment vectors for structures \textbf{1}(left) (1.252,-0.950), \textbf{2}(left) (-1.489,0.672), \textbf{3}(Tr1) (-1.652,0.356), and \textbf{3}(Tr2) (-1.042, 1.174) (in Debye).}
\label{fig:dipole}     
\end{figure}

For the interaction with the laser field we assume that the permanent dipole moment depends  only linearly on the coordinates, i.e.,
\begin{equation}
\label{eq:field}
H_{\rm field}(t)= -(d_a x_a + d_s x_s ) E(t)
\end{equation}
where $d_{a/s}$ is the derivative of the dipole moment with respect to $x_{a/s}$.  The permanent dipole moment changes mostly along one polarization direction ($x$, cf. Fig. \ref{fig:dipole}). Assuming the laser being linearly polarized along that direction and making a linear interpolation between the minimum configurations we obtain $d_s=-0.028$e and $d_a=0.150$e.

For the laser field we will use the following form
\begin{equation}
\label{eq:laser}
E(t)= \sum_{i=1,2} E_{0,i} \cos(\omega_i t) \exp(-(t-t_{0,i})^2/2\tau_i^2)
\end{equation}
which facilitates the consideration of the so-called pump-dump control mechanism \cite{korolkov96:10874,doslic98:292}. In Eq. (\ref{eq:laser}) $E_{0,i}$ is the amplitude, $\omega_i$ the carrier frequency, and $\tau_i$ the temporal width of the pulse centered at $t_{0,i}$.
%
%
\begin{table} [tp]%
\label{tab:S0Sn}
\centering
\caption{Excitation energies (in \cm) and oscillator strengths (in parenthesis) obtained at the TD-DFT B3LYP/6-31+G(d,p) level of theory. The calculated values for the different structures are compared with the experimental assignment from Refs. \cite{gil00:534} (\textbf{2},\textbf{3}) and  \cite{waluk91:5511} (\textbf{1}). Note that the experimental values correspond to the respective 0-0 transition.}
\begin{tabular}[t]{|l|cc|r|}
\hline
struct.                       &  calc.                    & exp.                 & leading excitation\\
\hline
\textbf{3}(Tr1)                     & 17392 (0.15) & 15708 &  144$\to$145(0.58)\\ 
\textbf{3}(Tr2)                     & 17670 (0.13) & 15993 &  143$\to$145(0.58)\\ 
\textbf{2}                            & 17536 (0.14) & 15983 &  128$\to$130(0.48)\\ 
\textbf{1}                            & 17904 (0.12) & 16000 &   80$\to$82(0.56)\\ 
\hline
\end{tabular}
\end{table}
%
\begin{figure*}[t]
\begin{center}
\resizebox{1.9\columnwidth}{!}{
\includegraphics{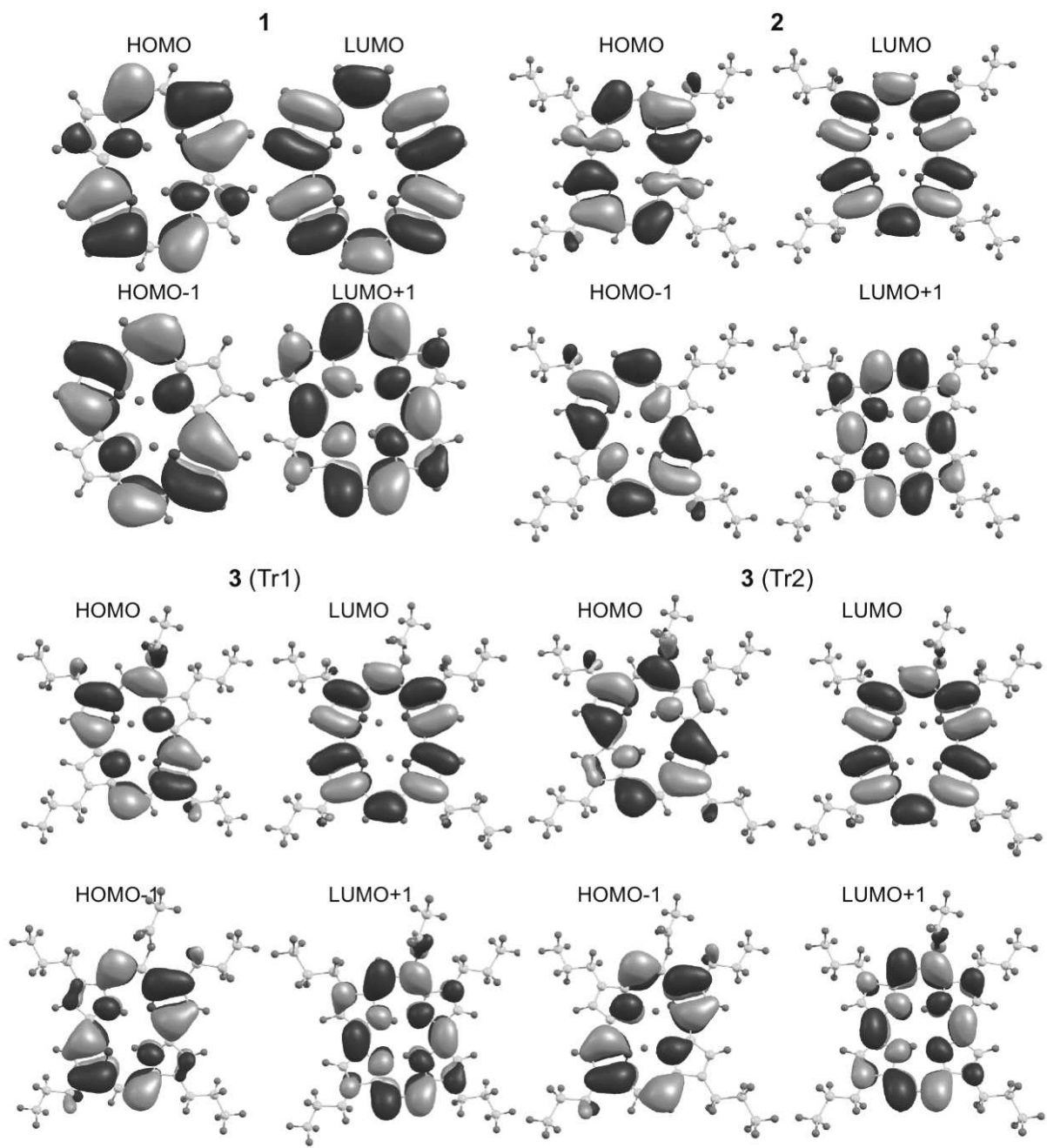}
}
\end{center}
\caption{Molecular orbitals relevant for the \Szo  transition (cf. Tab. 3) for molecules \textbf{1}, \textbf{2}, and the two tautomers of \textbf{3}. The orbital numbers are  80-83 (\textbf{1}), 128-131 (\textbf{2}), and 143-146 (\textbf{3}).}
\label{fig:MO}     
\end{figure*} 
%
%
\subsection{Molecular orbitals and excited states}
%
In the following we will discuss the \Szo {transition}, which contributes to the $Q$ absorption band.
Results of TDDFT calculations are compared with experimental data in Tab. 3 for the different compounds. First, we notice that TDDFT gives excitation energies which exceed the experimental ones by about 1500 - 2100 \cm. Most of this difference must be attributed to the method itself (see, e.g., systematic study in Ref. \cite{parac02:6844}). Further deviations might be due to the fact that the experimental values correspond to the 0-0 transitions whereas the calculations are performed for a vertical excitation. Comparing the experimental spectra of porphycences in gas and condensed phases the environmental effects (e.g., a nitrogen matrix in Ref. \cite{gil00:534}) is negligible \cite{waluk07:245}.

Let us focus on the tautomerization of \textbf{3}. As discussed in Section \ref{sec:geom} the energetic asymmetry in the electronic ground state is rather small. The measured splitting in the \Szo  transition, however, gave evidence for  a considerable asymmetry in the excited states of 285 \cms\cite{gil00:534}. The assignment of the most stable tautomer in the S$_1$ excited state has been based on a semiempirical ZINDO/S calculation. Considering Tab. 3 we notice that the agreement of the splitting in the \Szo tran\-sitions (278 \cm) is rather good thus giving strong support for the experimental conclusions in Ref. \cite{gil00:534}.

The \Szo transitions are of $\pi\rightarrow \pi^*$ type with participation of the $\pi$ (HOMO, HOMO-1) and $\pi^*$ (LUMO, LUMO+1) molecular orbitals (MOs). These orbitals are shown for the different molecules in Fig. \ref{fig:MO} and orbital coefficients are compiled in Tab. 3. 
The main contributions to the \Szo transition in \textbf{1} and \textbf{2}  is of HOMO-1 $\rightarrow$ LUMO type (see also Ref. \cite{sobolewski09:7714}). The same holds true for the less stable tautomer Tr2 of \textbf{3}. Interestingly, in tautomer Tr1 of \textbf{3} this transition is of HOMO $\rightarrow$ LUMO type. Comparison of the MOs of Tr1 and Tr2 shows
that the order of HOMO and HOMO-1 is \emph{reversed} upon tautomerization.

A closer look at the orbitals reveals the effect of symmetric and asymmetric substitutions. Going from \textbf{1} to \textbf{2} some electron density appears at those alkyl chains which are opposite to the position of the hydrogen bonded H atom. Overall, however, the MOs are rather similar, i.e. in the HOMO there is a $p$-orbital at those nitrogens, which are bonded to the central hydrogens whereas there is no electron density at the "free" nitrogen atoms. In the HOMO-1 this situation is reversed. Asymmetric substitution causes a disturbance of the porphycene $\pi$-system by the acetoxy group with the effect being different for the two tautomers. Tr2 of \textbf{3} resembles the HOMO in \textbf{1} and \textbf{2} with some electron density at the acetoxy group. In Tr1 of \textbf{3} the presence of the latter electron density renders the Tr2-HOMO-1 to become more stable than the Tr2-HOMO.

%
%
%
\subsection{Transition dipole orientation}
The directions of the transition dipole moments for the \Szo transitions of the different molecules are compiled in Fig. \ref{fig:dipole}. Comparing \textbf{1} and \textbf{2} we notice that symmetric substitution has almost no effect on the direction of the transition dipole moments. Furthermore, both cases show the well-known 90 degree change of the transition dipole moment upon tautomerization. The presence of the acetoxy group in \textbf{3}, however, changes this picture due to the polarity introduced by the oxygen atoms. Most notably is the fact that the tautomerization causes a tilt of the transition dipole moment by about 50 degrees only.

Let us assume that \textbf{3} is built into a molecular photonic wire. The F\"orster energy transfer coupling between different molecular subunits of the wire called donor (D) and acceptor (A) is given by  \cite{may10}
\begin{equation}
\label{ }
J_{DA}=\kappa_{DA}d_{10}^{(D)} d_{10}^{(A)} \frac{1}{R_{DA}^{3}}\, .
\end{equation}
Here, $d_{10}^{(A/D)}$ is the magnitude of the transition dipole moment for the \Szo transition and $R_{DA}$ is the distance between donor and acceptor. Most important for the present discussion is the  orientational factor given by
\begin{equation}
\label{eq:ft}
\kappa_{DA} = \mathbf{ n}_A\cdot \mathbf{ n}_B - 3(\mathbf{ e}_{DA}\cdot \mathbf{ n}_D)(\mathbf{ e}_{DA}\cdot \textbf{n}_A)\, ,
\end{equation}
where $\mathbf{ n}_{D/A}$ and $\mathbf{ e}_{DA}$ are the unit vectors pointing in the directions of the transition dipole moment of D/A and the center to center DA distance, respectively.

In order to estimate the effect of DPT on the F\"orster transfer coupling and in particular on the orientational factor we suppose that  \textbf{3} is built into the wire such that the neighboring unit has a transition dipole moment like structure \textbf{1}. This could be the case, for instance, for the widely used porphyrins. Let us further assume that $\mathbf{ e}_{DA}$ points along the $x$-direction. In this case we obtain $\kappa_{DA}$(Tr1)=1.43 and $\kappa_{DA}$(Tr2)=0.60. In other words the transfer coupling changes by about 60\% as a consequence of the tautomerization. 
%
\begin{figure}[h!]
\begin{center}
  \includegraphics[width=1.0\columnwidth]{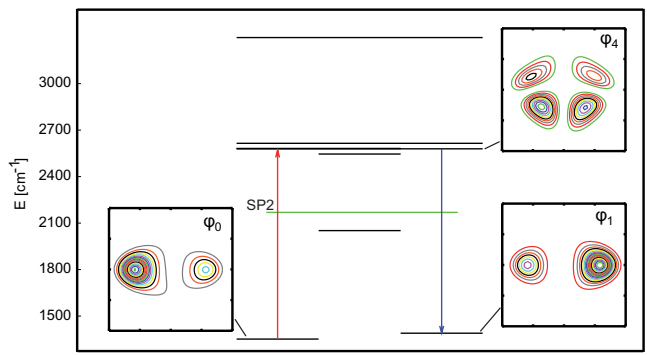}
\end{center}
\caption{Energy level scheme of the model Hamiltonian for \textbf{3} including the probability densities of those states which dominate the dynamics. The lengths of the horizontal lines indicate the area of delocalization of the wave function along the $x_s$ coordinate. The green line corresponds to the energy of the SP2 barrier. The vibrational eigenstates have been obtained using the improved relaxation scheme within MCTDH \cite{meyer06:179}. The pathway of laser-driven pump-dump control is marked by vertical arrows. The density plots cover the range $[-1:1]a_0$ along the vertical $x_a$ and the horizontal $x_s$ axes.}
\label{fig:levels}     
\end{figure} 
%
\subsection{Laser control of DPT in the electronic ground state}
\label{sec:control}
\begin{figure}[h!]
\begin{center}
 \includegraphics[width=0.85\columnwidth]{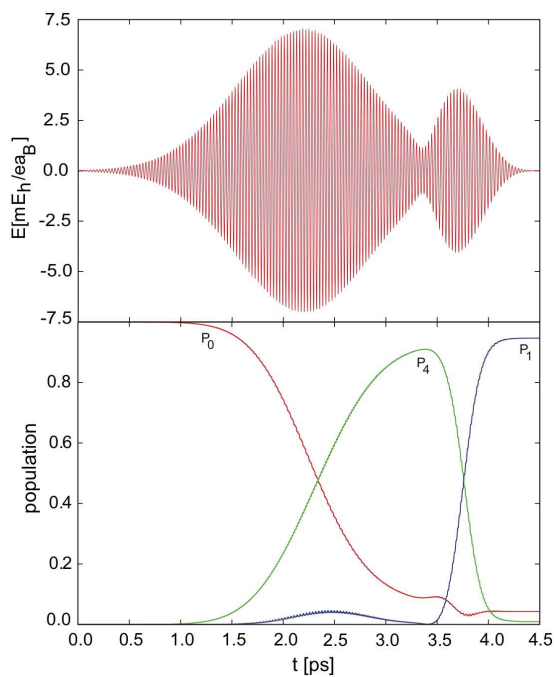}
\end{center}
\caption{Upper panel: Electric field for the pump-dump control scheme (pulse parameters: $E_{0,1}=7$ mE$_{\rm h}$/ea$_{\rm 0}$), $\omega_1=1111$ cm$^{-1}$, $t_{0,1}=2200$ fs, and $\tau_{1}=650$ fs;  $E_{0,2}=4$ mE$_{\rm h}$/ea$_{\rm 0}$), $\omega_2=1055$ cm$^{-1}$, $t_{0,2}=3750$ fs, and $\tau_{2}=200$ fs).
Lower panel: Population dynamics of the states  $\varphi_0$,   $\varphi_1$, and $\varphi_4$  }
\label{fig:pdv4}     
\end{figure}

In the following we will investigate the possibility to control the DPT in \textbf{3} using a simple two pulse pump-dump scheme involving an intermediate state which is energetically above the reaction barrier for concerted DPT (cf. Fig. \ref{fig:levels}). The initial state is localized mostly in the Tr1 well, while the final state lies mostly in the Tr2 well. Due to the low barrier and the small asymmetry both states have finite probabilities in the respective other wells. 
The goal is to find an electric field of the form given by Eq. (\ref{eq:laser}), for which  the first (pump) pulse populates the  intermediate state and the second (dump) pulse triggers a transition from this intermediate to the reactant state. This scheme operates in complete analogy to other isomerization reactions  studied previously, such as the single H-atom transfer in malonaldehyde \cite{doslic98:292} or the cope rearrangement in semibullvalenes \cite{korolkov96:10874}.
  The parameters for the present case of DPT were chosen as follows: First, the field strength is fixed at a certain value and the pulse length is increased starting from 800 fs until the intermediate state population reaches a maximum. For this pulse length the field amplitude is changed around its starting value to check whether the intermediate state population can be increased. The carrier frequency is chosen to correspond to the respective transition frequency. The same procedure is followed independently for the dump pulse by letting it start immediately after the pump pulse. This gave a total pulse duration of 5 ps during which the population of the intermediate state showed a plateau, which allowed us to shift the dump pulse such that pump and dump pulse partly overlap (cf. Fig. \ref{fig:pdv4}).

The progress of the laser-driven reaction can be  followed by calculating the populations of vibrational eigenstates of the model Hamiltonian. 
A representation of the energy level scheme is given in Fig. \ref{fig:levels} which also shows the probability density of those states which dominate the dynamics. The chosen above-barrier transition state involves an excitation of the asymmetric stretching coordinate $x_a$ which triggers the reaction to occur mostly along the stepwise pathway (see also Ref. \cite{abdel-latif10:76}). Furthermore, it is seen that the probability density has a higher weight in that part of the potential where $x_a<0$, implying that there is a preference for the Cis1 minimum to be visited during the dynamics. According to our previous study \cite{abdel-latif10:76} this is due to the higher stability of Cis1 as compared with Cis2. Finally, we note that there are several states in the region of the chosen intermediate state $\varphi_4$, thus state selective population will be more easily  achieved using long pulses as compared, e.g., with the NH-vibrational stretching frequency.

The optimized field and selected state populations are shown in Fig. \ref{fig:pdv4}. The  pump pulse excites the system from the ground state ground state, $\varphi_0$, to the  intermediate state $\varphi_4$, while the dump pulse de-excites this population to the target state $\varphi_1$. The maximum intermediate state population is 91 \%, reached around 3.4 ps. At this time, however, the dump pulse is already transferring population from $\varphi_4$ to the target state $\varphi_1$. The overall duration of the laser-induced DPT process is 4.5 ps and the final population of the target state is 95 \%. Two minor features seen in Fig. \ref{fig:pdv4} deserve a comment. Around 2.4 ps there is some transient population of the target state already due to the pump pulse. Later on around 3.6 ps the dump pulse also is seen to further depopulate the initial ground state. Both effects can be traced to the fact that the energetic separation between the two transitions is only 56 \cm, i.e. they can be simultaneously addressed by the laser pulse spectra. 
%
%
\section{Conclusions}
\label {sec:concl}
In summary, we have investigated the possibility to use IR laser-controlled double proton transfer in the electronic ground state of asymmetrically substituted porphycene as a way of influencing the direction of the \Szo transition dipole moment. This gives a means for changing the dipole-dipole interaction between neighboring chromophores in a molecular photonic wire. The estimated effect amounts to a 60\% change of the transfer coupling which could be used to design  a molecular switch operating on a time scale of a few picoseconds and being, in principle, reversible.

Needless to say that this proof-of-principle study calls for several extensions. First, the two-dimensional model of double proton transfer needs to be extended to account for heavy atom motions of the scaffold, e.g., by using the Cartesian reaction surface approach \cite{giese06:211}. Indeed Waluk and coworkers showed that the coupling to this type of motions may have a strong influence on the dynamics \cite{waluk06:945,waluk09:761,walewski10:2313}. It is to be expected that intramolecular vibrational energy redistribution is effective on the time scale of a few picoseconds covered in the present study, i.e. either substantially shorter pulses have to be used or more sophisticated pulse forms need to be devised, which, e.g., make use of the promoting character of certain scaffold modes. Second, a suitable design for a molecular photonic wire containing a DPT switching unit has to be found. This implies not only a favorable orientation of the transition dipoles of neighboring units but also the dominance of through space F\"orster transfer over through bond energy transfer.
\begin{acknowledgements}
This work has been in part financially supported by a scholarship  from the Ministry of Higher Education of the Arab Republic of Egypt.
\end{acknowledgements}



\end{document}